\documentclass[journal]{journal}
\usepackage[T1]{fontenc}

\usepackage{hyperref}
\usepackage{caption}
\usepackage{amsmath}
\usepackage{graphicx}
\usepackage{cite}
\usepackage{booktabs}
\usepackage{pifont}
\usepackage{graphicx}
\usepackage{listings}
\usepackage{chngcntr}
\usepackage{xcolor}

\usepackage{amsmath}
\usepackage{algorithm}
\usepackage{algorithmic}
\usepackage{tikz}

\makeatletter
\renewcommand{\ALG@name}{}
\makeatother

\usepackage{caption} % 使用caption宏包自定义标题格式

\DeclareCaptionFormat{listing}{#1#2#3} % 移除 \lstlistingname
\AtBeginDocument{%
}

\newcommand{\cmark}{\ding{51}} % 对号
\newcommand{\xmark}{\ding{55}} % 错号

\ifCLASSINFOpdf
\else
\fi
\begin{document}
\title{PVU: A Posit Vector Processor Unit Based on RISC-V Extension for Advanced Floating-Point Computation}
\author{Xinyu~Wu,
        Yaobin Wang${^\ast}$,
        Tianyi~Zhao,
        Jiawei~Qin,
        Zhu~Liang,
        and~Jie~Fu.% <-this % stops a space
\thanks{Wu, Zhao and Qin are students of the School of Computer Science and Technology at Southwest University of Science and Technology, Southwest University of Science and Technology, Mianyang, Sichuan, China, e-mail: cswuxy@mail.swust.edu.cn, zhaotianyi@mails.swust.edu.cn, yzgll@foxmail.com.}% <-this % stops a space
\thanks{Wang is a professor in the School of Computer Science and Technology at Southwest University of Science and Technology, Mianyang, Sichuan, China, e-mail: wangyaobin@foxmail.com.}% <-this % stops a space
\thanks{Liang and Fu are teachers in the School of Computer Science and Technology at Southwest University of Science and Technology, Mianyang, Sichuan, China, e-mail: liangzhu@mail.swust.edu.cn, fujie@mails.swust.edu.cn.}}


\markboth{Journal of \LaTeX\ Class Files,~Vol.~6, No.~1, January~2007}%
{Shell \MakeLowercase{\textit{et al.}}: Bare Demo of IEEEtran.cls for Journals}



\maketitle
\thispagestyle{empty}

\begin{abstract}
  %\boldmath
With the rapid development of edge computing, artificial intelligence and other fields, the accuracy and efficiency of floating-point computing have become increasingly crucial. However, the traditional IEEE 754 floating-point system faces bottlenecks in energy consumption and computing accuracy, which have become major constraints. To address this issue, the Posit digital system characterized by adaptive accuracy, broader dynamic range and low hardware consumption has been put forward. Despite its widespread adoption, the existing research mainly concentrates on scalar computation, which is insufficient to meet the requirements of large-scale parallel data processing. This paper proposes, for the first time, a Posit Vector Arithmetic Unit (PVU) designed using the Chisel language. It supports vector operations such as addition, subtraction, multiplication, division, and dot product, thereby overcoming the limitations of traditional scalar designs and integrating the RISC-V instruction extension. The contributions of this paper include the efficient implementation of the vector arithmetic unit, the parametric and modular hardware design as well as the verification of the practical application of the positive digital system. This paper extracts the quantized data of the first convolutional layer for verification. Experiments indicate that the accuracy rate of the division operation is 95.84\%, and the accuracy rate of the remaining operations is 100\%. Moreover, the PVU is implemented with only 65,407 LUTs. Therefore, PVU has great potential as a new-generation floating-point computing platform in various fields. 

\end{abstract}
  

  % \fi
  \IEEEpeerreviewmaketitle

  \section{Introduction}
% Here PVU have the typical use of a "T" for an initial drop letter
\IEEEPARstart{W}{ith} the continuous rise of Artificial Intelligence of Things (AIoT), the demands for numerical computation precision and computational efficiency in edge computing are constantly increasing. Although the traditional IEEE 754 floating-point system\cite{23} performs stably in most scenarios, its fixed precision allocation and rounding mechanisms have gradually revealed issues, such as a lack of flexibility, high energy consumption, and significant precision loss in certain specialized applications. To address these challenges, the Posit number system\cite{24}, proposed by John L. Gustafson in 2017, has gradually become one of the research hotspots in both academia and industry due to its advantages of adaptive precision distribution, wider dynamic range, and lower hardware resource consumption.

However, research on Posit floating-point arithmetic hardware is limited, and most studies adopt a scalar computation model, where data is processed sequentially. While this design is relatively simple and the implementation is more mature, it tends to create performance bottlenecks when handling large-scale parallel data processing. In contrast, vector arithmetic units, by performing parallel operations on multiple data elements simultaneously, significantly improve data throughput and computational efficiency. In fields such as signal processing, image processing, scientific computing, and machine learning, vector arithmetic units have demonstrated higher energy efficiency and lower power consumption, becoming a key breakthrough for solving complex computational tasks.RISC-V is an open, extensible architecture known for its simplicity, efficiency, and modularity, while the RISC-V Vector Extension (RVV) enhances performance by enabling parallel computations on multiple data elements, benefiting fields like scientific computing, image processing, and machine learning\cite{21}.
    
For the reasons mentioned above, this study aims to design and implement PVU based on the Chisel language\cite{22}. The unit supports operations such as addition, subtraction, multiplication, division, and vector dot products. Additionally, by utilizing customized RISC-V instruction extensions, it breaks the limitations of traditional scalar designs and provides a novel solution for next-generation high-performance, low-power computing platforms. Specifically, the main contributions of this work are as follows:
   
\begin{itemize}
  \item Parameterized and Modular Vector Arithmetic Unit: The proposed PVU can efficiently and parallel - process multiple datasets. It supports five operations: addition, subtraction, multiplication, division, and dot - product. Moreover, it enables parameterized configuration of bit - width, exponent bit - width, and mantissa alignment bit - width. This effectively addresses the performance bottleneck issues encountered in large - scale data processing and the limitation of only supporting specific bit - widths in existing Posit arithmetic hardware. 
  \item RISC-V ISA Customization for Posit-Based Vector Processing Acceleration: This work presents a co-design architecture. This architecture realizes Posit vector operations through customized RVV instructions, providing software support at the ISA level to directly call the Posit vector arithmetic unit. This architectural innovation effectively bridges the long - standing abstraction gap between software and hardware in the Posit computing system.
  \item Posit Number System Application Verification: Experiments indicate that the accuracy rate of the division operation is 95.84\%, and the accuracy rate of the remaining operations is 100\%. Moreover, the PVU is implemented with only 65,407 LUTs and 108 Muxes.It has been demonstrated that PVU can achieve high-precision floating-point calculations with relatively low overhead, breaking through the computational bottleneck of IEEE 754.
\end{itemize}  

The structure of the remaining parts of this paper is as follows: Section 2 introduces the Posit number system and the RISC-V instruction set. Section 3 compares the similarities and differences between this study and other existing research. Section 4 provides a detailed description of the logical implementation of the PVU, with a focus on the multiplication module. Section 5 discusses the custom instruction set architecture extensions and compilation support. Section 6 describes the hardware cost and computational precision. The final section presents the conclusion.

\section{Background}
The Posit number system improves the efficiency and precision of numerical representation by introducing a flexible encoding mechanism. As shown in Fig.~\ref{fig1}, its basic components include  the Sign, Regime, Exponent, and Fraction. Each component has a unique function and structure, allowing Posit to achieve a balance between precision and range. The following is a detailed introduction to each component:

\begin{figure}
    \includegraphics[width=3.5in]{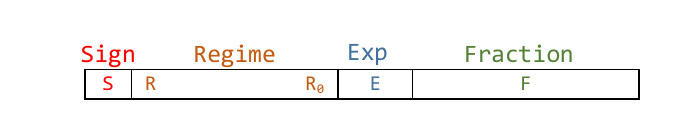}
    \caption{The posit number system format} \label{fig1}
    \end{figure}

The Sign bit (S) is the first bit in the Posit number system and is responsible for determining the sign of the Posit number. If \( s = 0 \), Posit is positive; if \( s = 1 \), Posit is negative.

The Regime field (R) is one of the key innovations of Posit. It uses a variable-length encoding scheme to represent the range of the exponent, determining the scale of the Posit. The Regime is composed of consecutive 1s and 0s, with the length controlled by the inversion bit \( R_0 \). When the Posit is large, the Regime will contain more 1s; conversely, when the Posit is smaller, the Regime will contain more 0s. The length of the Regime field is \( k \), and the value \( r \) is calculated using the following formula:

\begin{equation}
    r =
    \begin{cases}
    k-1, & \text{if } R_0 = 1 \\
    -k, & \text{if } R_0 = 0
    \end{cases}
\end{equation}

The Regime value is a special constant scaling factor \( U_{\text{seed}} \), and its value depends on the configuration of the Posit. Specifically, it is determined by the bit width \( E_S \) of the Exponent part in the Posit. The calculation formula is as follows:

\begin{equation}Useed=2^{2^{es}}\end{equation}

The Exponent field (E) is the part of the Posit that represents the configured exponent. Since the Regime field has a variable length, the Exponent bits may appear after the least significant bit (LSB) of the Posit, in which case the value of the Exponent will be 0.

The Fraction field (F) is the final part of the Posit number system, similar to the mantissa part in IEEE 754. However, there is always an implicit number \( m \) in front of the Fraction. When the Posit is positive, \( m = 1 \); when the Posit is negative, \( m = 2 \). This means that in actual storage, the mantissa part typically stores only the subsequent binary digits, with the first bit not needing to be explicitly stored. For example, the value \( 1.101 \) will store the bits \( 101 \) in the Fraction part, implicitly understood as \( 1.101 \).

Combining the four components mentioned above, although in the Posit Standard 2022, \( E_S \) is fixed at 2, for better evaluation, the PVU parameterizes \( E_S \), resulting in two calculation formulas. When \( E_S \neq 2 \), the Posit value \( p \) is calculated as shown in Equation (3), and when \( E_S = 2 \), the Posit value \( p \) is calculated as shown in Equation (4).

\begin{equation}{p}=((1-3s)+f)\times2^{(1-2s)\times(e+s)}\times Useed^r\end{equation}

\begin{equation}{p=((1-3s)+f)\times2^{(1-2s)\times(4r\times e+s)}}\end{equation}

Specially, if the entire Posit file consists of 0s, the Posit value is 0. If the entire Posit file consists of 1s, the Posit value is Not a Real (NaR), which is the umbrella value for anything not mathematically definable as a unique real number.

\begin{figure}
    \includegraphics[width=3.5in]{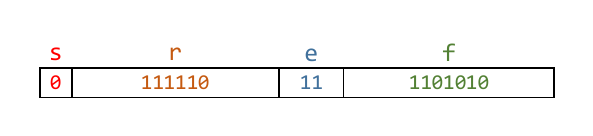}
    \caption{The decoding example of Posit<16, 2>} \label{fig2}
\end{figure}

\begin{table*}
  \caption{The similarities and differences between this work and related Posit processing units}\label{table1}
  \resizebox{\textwidth}{!}{%
  \begin{tabular}{|l|c|c|c|c|c|c|c|}
  \hline
   & THIS WORK & BIG-PERCIVAL & PERC & PERI & POSAR & PPU-Light & PDPU \\
  \hline
  Configurable posit size & \cmark & \cmark & \cmark & \cmark & \cmark & \cmark & \cmark \\
  Basic arithmetic        & \cmark & \cmark & \cmark & \cmark & \cmark & \xmark & \xmark \\
  Dot product / FMA       & \cmark & \cmark & \cmark & \cmark & \xmark & \xmark & \cmark \\
  RISC-V integration      & \cmark & \cmark & \cmark & \cmark & \cmark & \cmark & \xmark \\
  Scalar operation        & \cmark & \cmark & \cmark & \cmark & \cmark & \cmark & \cmark \\
  Vector operation        & \cmark & \xmark & \xmark & \xmark & \xmark & \xmark & \xmark \\
  Advanced software support & \cmark & \xmark & \xmark & \xmark & \xmark & \xmark & \cmark \\
  Custom instruction support & \cmark & \cmark & \xmark & \cmark & \xmark & \cmark & \xmark \\
  Open source             & \cmark & \cmark & \xmark & \xmark & \cmark & \cmark & \cmark \\
  \hline
  \end{tabular}%
  }
\end{table*}

For example, as shown in Fig.~\ref{fig2}, \( 0111110111101010 \) is the binary encoding of Posit<16,2>. The Sign bit of the Posit is 0, indicating that it is positive, so \( s = 0 \). The Regime field is \( 111110 \), consisting of 5 consecutive 1s and 1 0, with a length \( k = 6 \), thus \( r = 5 - 1 = 4 \). The Exponent field is \( 11 \), so \( e = 3 \), and the \( U_{\text{seed}} \) value is 16. Finally, the Fraction field is \( 1101010 \), so \( f = \frac{106}{2^7} = \frac{106}{128} = 0.828125 \). Therefore, using the formula (3), so \( p = (1 + 0.828125) \times 2^{(1 - 2 \times 0)} \times (4 \times 4 + 2 + 0) = 239616 \).

\section{Related work}

Regarding the RISC-V architecture, the European PULP platform has explored low-power applications of RISC-V, particularly excelling in Internet of Things (IoT) devices and embedded systems.\cite{1} The advantage of RISC-V lies in its simple and highly modular instruction set, which can be extended according to specific needs, making it especially suitable for implementing customized designs on edge devices, such as accelerators dedicated to artificial intelligence inference tasks\cite{2}.

In the field of Posit floating-point software, SoftPosit is an open-source project developed by Berkeley Lab\cite{3}, which provides basic arithmetic support for Posit floating-point numbers and demonstrates the potential applications of Posit in high-precision numerical computations. The Posit library for Julia\cite{4} has implemented support for Posit floating-point numbers and is widely used in numerical computing and deep learning inference tasks. Research indicates that, under the same bit-width, Posit floating-point numbers significantly outperform IEEE 754 floating-point numbers in terms of computational precision and efficiency\cite{5}, especially in low-power, high-efficiency computing scenarios, where Posit has broad application prospects. There are also numerous studies in other fields that apply the Posit format, including image processing\cite{6}, weather forecasting\cite{7}, and more.

The team led by Florent de Dinechin at Lyon University in France explored the applicability of Posit floating-point numbers\cite{8} and found them to be highly suitable for the field of machine learning, demonstrating an exact conversion between Posit and IEEE 754 floating-point numbers. The team at Rochester Institute of Technology in the United States, led by Zachariah Carmichael, successfully ran deep neural networks (DNNs) on Posit floating-point numbers with less than 8 bits\cite{9}, achieving better accuracy and lower latency. The team at the University of Pisa in Italy, led by Marco Cococcioni, explored the potential for vectorized computation with Posit floating-point mechanisms through software simulation, integrating this into the DNN model training and inference process\cite{10}. The PeNSieve project at the Complutense University of Madrid proposed a Posit-based framework for training and inference of deep neural network models, incorporating low-bit quantization through fused operations\cite{11}.

Compared to the widespread software simulation and application of the Posit number system, research on hardware computation units is still relatively scarce. The PERCIVAL\cite{12} and BIG-PERCIVAL\cite{13} projects at Complutense University of Madrid focus on the implementation of small-scale and large-scale Posit computations, aiming to accelerate deep learning inference and scientific computing. The PERC project\cite{14} at Aachen University of Technology in Germany studies how to integrate Posit operations into high-performance processors, while the PERI project\cite{15} at the Indian Institute of Technology Madras focuses on optimizing high-precision computation and energy efficiency for IoT devices. The POSAR project\cite{16} at the National University of Singapore designs lightweight Posit processing units for smart terminal devices and low-power applications, driving the widespread use of the RISC-V architecture in edge computing. The PPU-light project\cite{17} at the University of Pisa, Italy, enables the conversion of Posit numbers into other numeral systems. The team led by Li Qiong at Nanjing University has designed a configurable open-source Posit Dot Product Unit (PDPU) that can perform efficient dot product operations in deep learning applications and supports mixed precision\cite{18}. This paper summarize some of these works and compare them with this study in TABLE.~\ref{table1}, PVU present first vector Posit computing system that integrates both basic arithmetic and fundamental linear algebra operations. This system supports both scalar and vector computations. By building upon RISC-V integration, it achieves enhanced high-level software support through inline assembly.

\section{Posit Vector Unit Arithmetic Operations And Implementation}
\begin{figure*}
    \includegraphics[width=\textwidth]{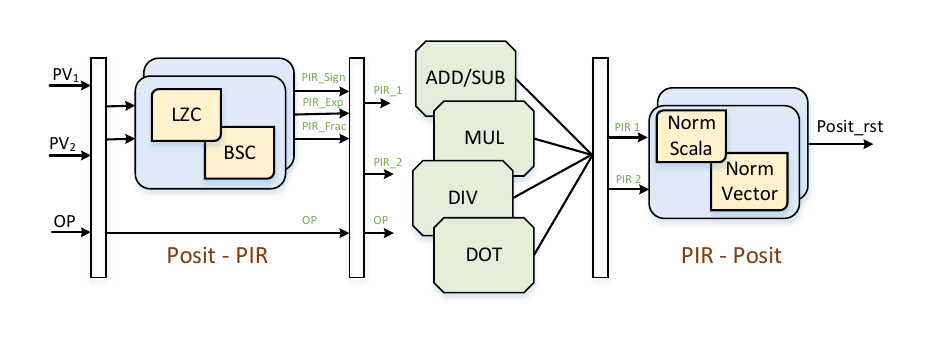}
    \caption{The structure of the posit vector processor unit} \label{fig3}
    \end{figure*}

The architecture of the PVU is shown in Fig.~\ref{fig3}. The PVU takes two Posit vector operands ($PV_{1}, PV_{2}$) and an operation mode number ($PV$) as inputs and outputs the corresponding Posit operation result. PVU have parameterized the design of the PVU, including the Posit bit width, Exponent bit width, and alignment width.

\subsection{Decode}

After the Posit operands enter the PVU, they are first subject to vector decoding. As shown in ~\ref{logic1}. First, this paper extract the sign bit. If the sign bit is 1, the Posit is negative; if the sign bit is 0, the Posit is positive. The extracted Posit binary is then modified to its two's complement representation.

Next, calculate the bit width and value of the Regime. The value of the Regime part is computed using a Leading Zero Count (LZC) module. First, PVU extract the highest bit of the Regime part. If it is 0, the Regime is of the form (00....01), and PVU directly pass it to the LZC module for counting. If it is 1, the Regime is of the form (11....10), and  first invert all the binary digits before passing it to the LZC module for counting. The LZC module will return the number of leading zeros and whether it is all zeros. The former can be used with the highest bit of the Regime to substitute into Equation (1) to calculate the value of \( r \). Since the length of the Regime field is dynamic, there is a case where the Regime field occupies the entire Posit bit width. In this case, it can be determined using the latter condition.

The length of the Regime field, \( k \), can be obtained by adding 1 to the number of leading zeros. then left-shift the Posit by \( k+1 \) bits using a barrel shifter, which allows us to extract the Exponent field from the highest bits. The bit width of the Exponent, \( E_S \), is determined by the Posit configuration (typically 2). If \( E_S \neq 2 \), PVU can substitute the \( E_S \) value into Equation (2) to calculate the \( U_{\text{seed}} \) value. If \( E_S = 2 \), no calculation of \( U_{\text{seed}} \) is needed. For ease of subsequent calculations, combine the Regime and Exponent to calculate a unified binary exponent value, \( \text{exp} \). Specifically, left-shift \( r \) by \( E_S \) bits and add the value of the Exponent field \( e \), yielding the value of \( \text{exp} \).

Lastly, left-shift the Posit by \( E_S \) bits using a barrel shifter to obtain the Fraction part. Note that in the Posit number system, the Fraction part has an implicit bit \( m \). PVU concatenate this implicit bit with the extracted Fraction to obtain the actual mantissa value used in the computation. At this point, PVU have successfully decoded the Posit and obtained the intermediate representation for Posit Intermediate Representation (PIR), which includes the sign indicating whether the Posit is positive or negative, the binary exponent value \( \text{exp} \), and the actual mantissa \( \text{frc} \).

\begin{figure*}
  \centering
  \includegraphics[width=\textwidth]{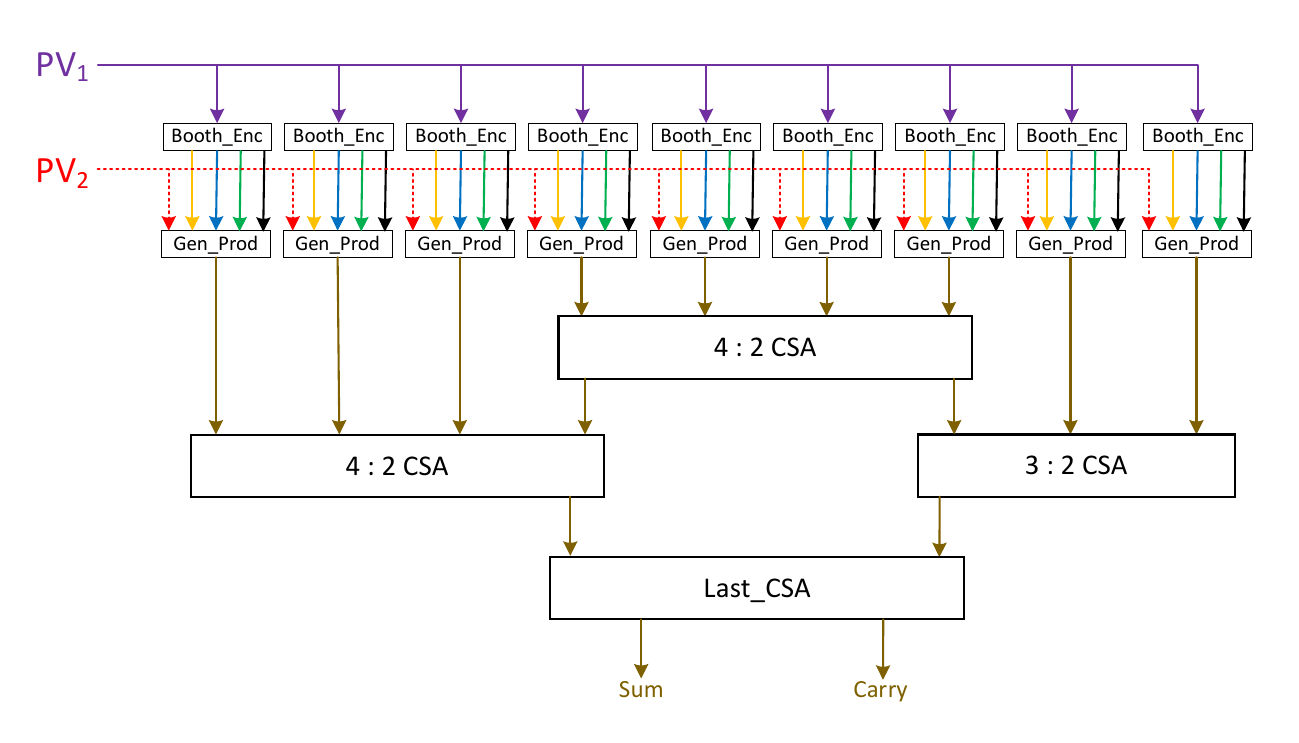}
  \caption{The design examples of Radix-4 Booth multiplier and CSA} \label{fig4}
\end{figure*}

\begin{algorithm}
  \caption{Posit Decode}
  \begin{algorithmic}[1]
  \label{logic1}
  \STATE \textbf{Input:} Posit operand $P$

  \vspace{0.5em}

  \IF{$s = 1$}
      \STATE $P \gets \text{negative}$
  \ELSE
      \STATE $P \gets \text{positive}$
  \ENDIF
  
  \vspace{0.5em}

  \IF{Regime\_MSB = 0}
      \STATE $r \gets 00\ldots01$
  \ELSE
      \STATE $r \gets 11\ldots10$
  \ENDIF
  
  \vspace{0.5em}
  
  \STATE $\textbf{LZC} \gets P$ 

  \STATE $k \gets \textbf{LZC}_{\text{out}} + 1$
  
  \vspace{0.5em}

  \IF {$E_S$ = 2}
      \STATE $U_{\text{seed}} \gets Equation (4)$
  \ELSE
      \STATE $U_{\text{seed}} \gets Equation (3)$
  \ENDIF

  \vspace{0.5em}

  \STATE $ \textbf{PIR}_{sign} \gets s$
  \STATE $ \textbf{PIR}_{exp} \gets r \ll E_S \; | \; e$
  \STATE $ \textbf{PIR}_{frac} \gets implicit.f$
  
  \STATE \textbf{Output:} $\text{PIF}_{sign}$, $\text{PIR}_{exp}$, $\text{PIR}_{frac}$
  
  \end{algorithmic}
\end{algorithm}

\subsection{Add/Sub}
After decoding, the computation phase begins, and all operations are performed based on PIR. Before performing addition and subtraction, the operands need to be aligned so that the exponents of the two operands are unified. Specifically, PVU first use a comparator to find the maximum exponent value, which will serve as the target value for alignment. then calculate the difference between each exponent and the target, modify the exponent values, and shift the mantissa based on the exponent difference, ensuring that the actual mantissa value remains unchanged. It is important to note that the number of bits to be shifted cannot exceed the alignment value configured for the Posit, to ensure that the precision of the mantissa is not lost during the computation.

After the alignment is completed, the corresponding operation is performed based on the OP. Since the addition and subtraction operations share similar logic, PVU will introduce them together here. First, PVU need to check the sign bits of the two operands to determine their signs. Next, since the exponents of the two aligned operands are the same, the exponent result of the addition/subtraction operation is the exponent target from the alignment process. At this point, only the mantissas need to be added or subtracted, and the corresponding carry/borrow is recorded. Finally, the final sign is computed based on the carry/borrow values and the sign bits of the operands. At this stage, the sign, exponent, and mantissa values of the PIR are all calculated.

\subsection{Mul}
The vector multiplication module is one of the core components of the PVU, and here PVU will focus on the implementation logic of this module. For multiplication, PVU do not need to perform the alignment process on the operands as PVU do for addition and subtraction. The multiplication module can directly process the decoded PIR vectors. First, the calculation of the sign bit is very simple; PVU can obtain the final sign bit by performing an XOR operation between the sign bits of the two operands.

Then, compute the mantissa part, which is the most complex section of this module. To better utilize hardware parallelism and achieve faster and more accurate mantissa multiplication, PVU adopt the base-4 Booth multiplication algorithm in the design. Fig.~\ref{fig4} illustrates this design logic. This algorithm effectively reduces the number of additions required for the multiplication operation by decomposing the multiplier into a base-4 encoded form. This results in higher speed and lower energy consumption, especially when handling large-scale data, and thus improves multiplication efficiency in hardware implementations compared to traditional multiplication algorithms. Specifically, PVU first pad the multiplier with zeros to align it to the format required by the signed Booth encoding, then perform the base-4 Booth encoding operation. The result is stored in the vector \texttt{codes}. Next, PVU decode the elements in \texttt{codes} sequentially, and according to the base-4 Booth multiplication algorithm, perform the appropriate calculations (such as bit inversion and left shifts, multiplication by 0, etc.). After the computation, the shifted results are concatenated to form a partial product.

Now, PVU need to perform the accumulation of the partial products. To improve the accumulation speed, PVU have designed a carry-save adder (CSA) tree unit using 4:2 compressors and 3:2 compressors. By utilizing a divide-and-conquer structure and recursive methods, PVU reduce the delay caused by carry propagation, thereby enhancing the overall performance of the addition process and efficiently handling large-scale accumulation operations. PVU input the partial products into the CSA, and the sum and carry are obtained. At this point, the implementation of the base-4 Booth multiplication algorithm is fully completed.

Lastly, perform the final summation of the sum and carry to complete the mantissa multiplication. The exponent calculation is much simpler; PVU only need to add the exponents of the two mantissas. It is worth noting that an overflow might occur after the summation, so PVU have set a maximum exponent to avoid this issue.

\subsection{Div}
The difficulty in floating-point vector division lies in the handling of mantissa division. For the sign and exponent, the handling is the same as multiplication: the sign is determined through XOR computation. The exponent, on the other hand, can be calculated through integer subtraction.

\begin{equation}\frac{1.1101...1011101}{1.0010...1000011}=\frac{11101...1011101\times2^{-F}}{10010...1000011\times2^{-F}}\end{equation}

In the mantissa division section, as shown in Equation (5), convert the division operation into a multiplication operation and reuse the base-4 Booth multiplier for multiplication. At this point, the key issue is how to convert the integer into its corresponding reciprocal. To address this, PVU use Newton's method to compute the reciprocal, with the iteration formula shown in Equation (6). Here, \( X_n \) is the current approximation of the reciprocal, and \( \text{num} \) is the input integer. Through multiple iterations, \( X_n \) will converge to the approximation of \( \frac{1}{\text{num}} \). The more iterations performed, the higher the precision, but also the greater the delay. PVU set the number of iterations to 3.

\begin{equation}X_{n+1}=X_n\times\left(2-num\times X_n\right)\end{equation}

Finally, PVU take the sum of the result and carry outputted from the base-4 Booth multiplier, and perform a shift operation to scale it back to the fixed-point format bit width.

\subsection{Dot Product}
In deep learning, the frequency of dot product operations is very high, making it essential to support a dedicated dot product computation module in the PVU. As shown in Equation (6), PVU decompose the dot product calculation into two parts. The first part is the vector multiplication operation, where the result is stored in an intermediate vector. The second part is the accumulation operation, where the elements of the intermediate vector are summed to obtain the final dot product result.

First, PVU reuse the existing vector multiplication module, and store the result in an intermediate variable. Before performing the accumulation, PVU need to align all the elements within the intermediate variable. In the addition/subtraction module, the alignment operation is performed for two operands as a pair, while in dot product operations, the alignment must be applied to all elements within the intermediate vector. Therefore, PVU have developed an alignment module specifically adapted for dot product operations, with an internal implementation logic identical to that of the addition/subtraction alignment module.

After the alignment is completed, PVU convert all the mantissas to two's complement format based on the sign bits, and reuse the CSA for accumulation. Since the CSA operates with a larger bit width during computation and performs a rounding operation after all calculations are completed, this significantly improves the precision of the dot product computation.

\subsection{Standardization}
As mentioned earlier, the Posit mantissa has an implicit bit. Before encoding the final result into Posit format, PVU need to normalize all the mantissas in the PIR. PVU treat the highest bit of the actual mantissa as the implicit bit and perform normalization to ensure that this bit is always 1.

PVU first invoke the LZC module to calculate the number of leading zeros in the mantissa. Then, using the number of leading zeros and the configured decimal point position, PVU calculate the adjustment required for the exponent. Finally, PVU use a barrel shifter to adjust the exponent so that the value before the decimal point is 1.

Lastly, PVU save the mantissa bit width plus 1 bit, perform RNE rounding on the lower bits, and adjust the exponent values in the PIR one by one. It is important to note that the output of the dot product operation is a scalar, while the output of other operations is a vector. Therefore, the normalization module has both scalar and vector versions, but the internal logic is the same.

\subsection{Encode}
After the computation is completed, PVU need to encode the various parts of the PIR into the specified Posit format. In fact, encoding is the inverse process of decoding, and PVU can handle it using the reverse process of decoding. The challenge lies in extracting the Regime part, as its dynamic bit width property means that the bit width of the Regime part after computation is unknown. Therefore, PVU first extract the Exponent part, which has a fixed bit width, and then extract the Regime part. In the Regime value computation part, to maintain the inversion bit property, PVU initialize the Regime value as 1 (00...01), then determine whether to invert it based on its sign. Finally, PVU calculate the Regime bit width inversely using the formula (1), completing the encoding of the Regime part.

Finally, PVU perform RNE rounding on the mantissa, concatenate the three parts, and convert them into two's complement representation to complete the Posit encoding. Note that since the result of the dot product operation is a scalar, PVU have also written a scalar encoding unit, which shares the same internal logic as the vector encoding unit.

\section{RISC-V ISA Extension And Compiler support}
At the instruction set architecture level, PVU designed the necessary vector instructions for the PVU using the RVV extension. As shown in TABLE.~\ref{table2}, PVU have customized corresponding instructions for the operations ADD, SUB, MUL, DIV, and DOT. Since the Posit number system is intended to replace the traditional IEEE 754 floating-point type, the custom instruction format follows the OPFVV instruction format. PVU use the custom encoding space \( 001101 \) to replace the original func6 field to indicate that the instruction is for Posit operations, and differentiate the various Posit operations using the fun3 field. For the vector mask, PVU simplified \( V_m \) to 1, indicating the use of a vector mask.

To support the use of custom instructions by hardware, PVU wrote a set of functions using inline assembly to map high-level C/C++ function calls to low-level machine code. As shown in Listing~\ref{lst:ass} and Listing~\ref{lst:vmul}, PVU do not need to modify any RISC-V compilers, as the correct data flow for Posit instructions is generated during the compilation process.
    
\begin{lstlisting}[caption={Posit inline assembly example}, label={lst:ass}]
posit_t* vmul(long* a, long* b, long* result, int len) {
    register long* p1 asm("v0") = a;  
    register long* p2 asm("v1") = b;  
    register long* res asm("v2") = result;  
    register int length asm("v3") = len;  
    __asm__(
        ".set vlen, 4\n"  
        ".set vli t0, %0, e32, m4\n"  
        ".set op, 0x57\n"                
        ".set opf1, 0x2\n"               
        ".set opf2, 0x0d\n"              
        ".byte op | ((r%[vd] & 1) << 7), "   
        "((r%[vs2] & 0xF) << 4) | ((r%[vs1] >> 4) & 0xF), " 
        "((r%[vs1] >> 8) & 0x1) | (opf1 << 1), " 
        "(op << 1)\n"                  
        : [vd] "=r"(vd)                  
        : "r"(vs1), "r"(vs2), "[vd]"(vd) 
    );
    return 0;
  }
\end{lstlisting}

    \begin{lstlisting}[caption={Posit vector convolution example},label={lst:vmul}]
  void conv4x4_vectorized(posit_t *a,  
    posit_t *f,  
    posit_t *c,  
    int n) {  
      for (int i = 0; i < n; i++) {  
       for (int j = 0; j < n; j++) {  
        posit_t sum = 0;  
         for (int k = 0; k < 4; k++) {  
          for (int l = 0; l < 4; l++) {  
           vect_t input_row = load_vector(&a[(i + k) * n + j], 4); 
           vect_t filter_row = load_vector(&f[k * 4 + l], 4);
           vect_t result = pmulvv(input_row, filter_row); 
           sum = padd(sum, sum_vector(result)); 
           }  
         }  
           c[i * n + j] = sum;  
        }  
      }  
    }\end{lstlisting}

An example of inline assembly for vector multiplication is shown in Listing 1. Here, PVU assume each element is 32 bits, and each operation processes 4 elements at a time. The register keyword allows the compiler to place the input and output vectors in the specified registers and set the vector length. In the inline assembly, lines 9, 10, and 11 are where PVU set the OP opcode, funct3 field, and funct6 field for specific Posit operations, which are the key parts distinguishing them from other instructions and operations.

Convolution is a commonly used operation in deep learning. In Listing 2, PVU present an example algorithm for performing a 4\xmark4 convolution using the PVU. The convolution kernel is vectorized by rows, with both the rows and columns loaded into the computation vector in one step. The result is computed all at once using vector multiplication and stored in the result vector. Finally, the result is summed using vector addition to complete the convolution computation.

\begin{table}[htbp]
  \centering
  \caption{\centering Custom posit operation instructions based on the RISC-V V extension}
  \includegraphics[width=3.5in]{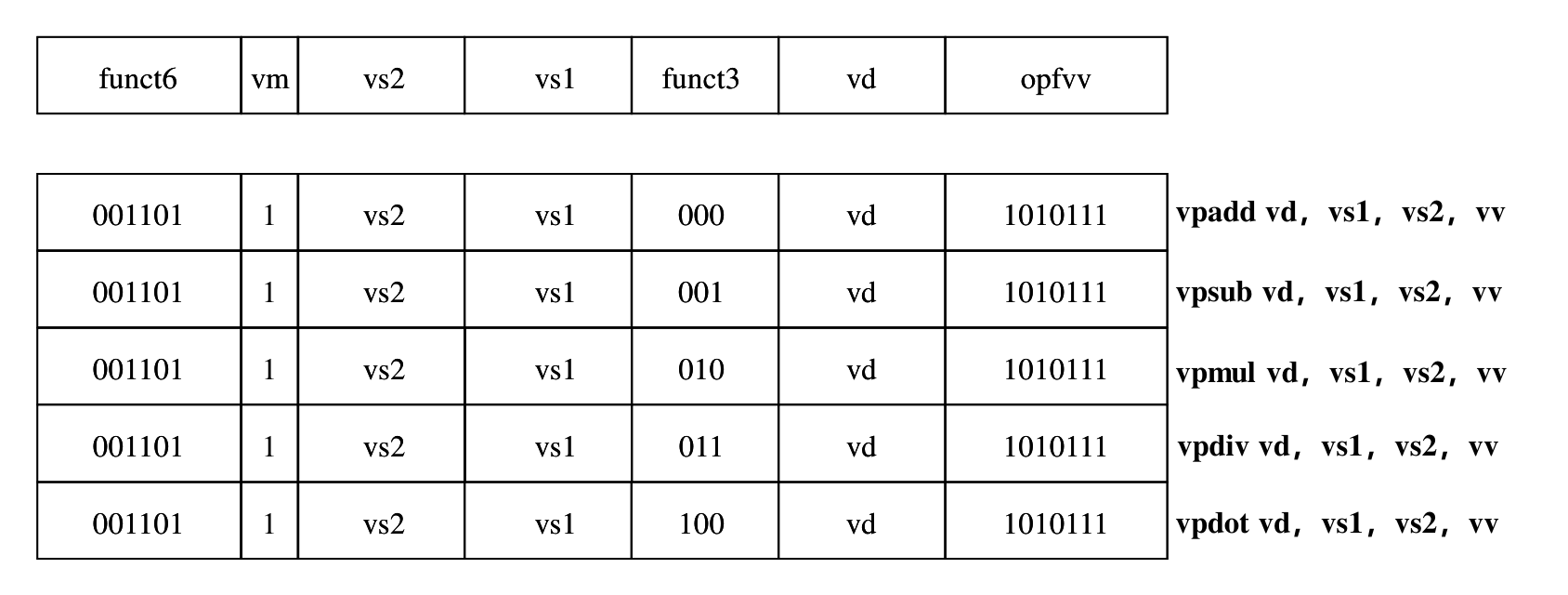}
  \captionsetup{justification=centering}
  \label{table2}
\end{table}

\section{Experimental Results}
To evaluate the computational correctness of each operation module in the PVU, PVU divided Posit into vectors as shown in TABLE.~\ref{table3}, and extracted the quantized activation values and weights from the first convolution layer of ResNet-18\cite{19} using PyTorch, saving them in FP64 format. Since the PVU operates using the Posit number system, PVU used the SoftPosit library to batch convert the extracted data to Posit32 and perform addition, subtraction, multiplication, division, and dot product calculations. This framework provided the necessary inputs and outputs for testing. The Kconfig framework was developed and tests were conducted, where the PVU achieved 100\% accuracy in vector addition, vector subtraction, vector multiplication, and vector dot product operations. The accuracy for vector division was 95.84\%, attributed to errors introduced by the reciprocal conversion.

\begin{table}[htbp]
    \centering
    \caption{\centering Examples of vector partitioning of each part of Posit}
    \includegraphics[width=3.5in]{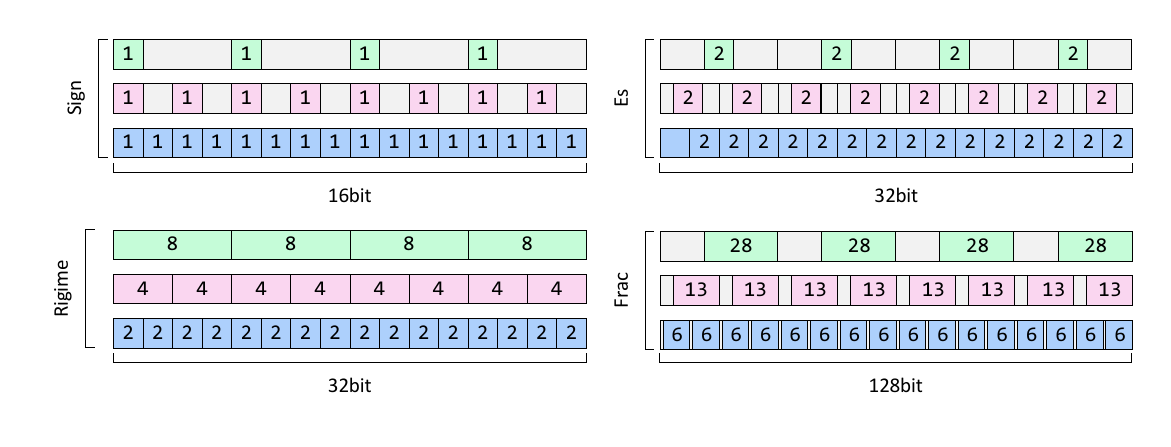}
    \label{table3}
\end{table}
  
In addition, PVU tested the accuracy difference between the Posit number system and FP32 in DNNs using Deep PeNSieve\cite{20}. As shown in Figures 6 and 7, PVU conducted tests on MNIST, Fashion-MNIST, SVHN, and CIFAR-10 at the TOP-1 and TOP-5 levels. For MNIST, the performance of Posit16 is on par with that of FP32, while Posit32 is slightly inferior to FP32. For Fashion - MNIST and SVHN, both Posit16 and Posit32 outperform FP32. Regarding CIFAR-10, the advantages of Posit become prominent, especially for Posit16. It has an accuracy 4.5\% higher than FP32 in TOP-1 and 2.4\% higher in TOP-5.It can be observed that Posit16 achieves higher accuracy than FP32 while reducing memory usage by half.

FPGA tests on the PVU were also conducted through Vavido. As shown in TABLE.~\ref{table4}, PVU have statistically analyzed the overall hardware consumption of PVU and the hardware consumption of the multiplication, division, and dot product operation components, because these three operations have the highest 
resource occupancy rates. The test results indicate that the overall overhead of PVU is acceptable. 

\begin{figure}[htbp]
  \centering
  \begin{minipage}[b]{0.45\textwidth}
    \centering
    \includegraphics[width=\textwidth]{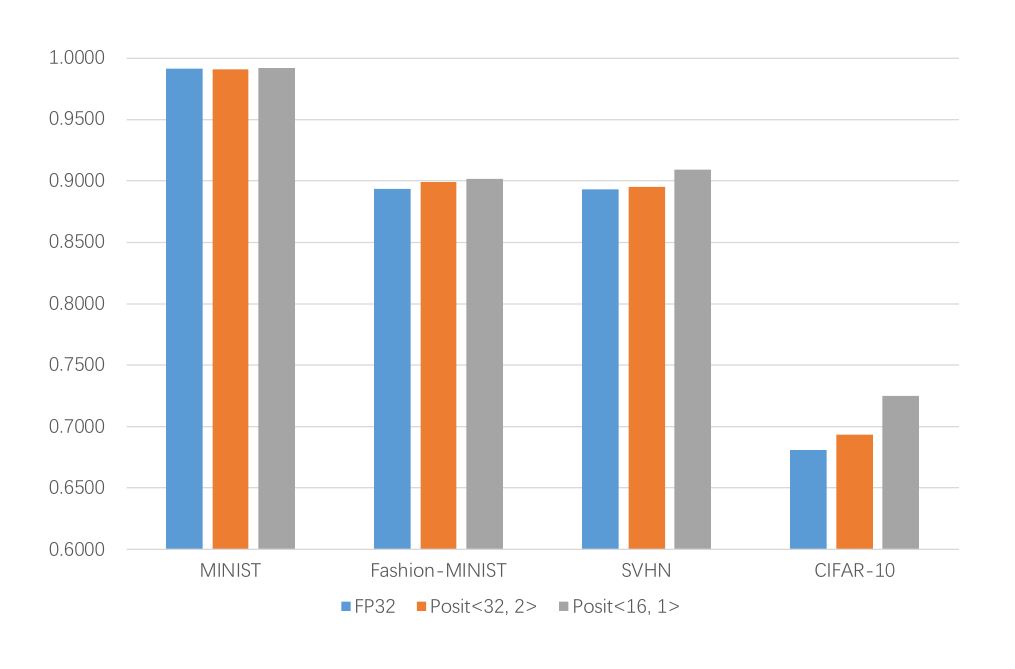}
    \caption{The accuracy of Posit-FP32 in TOP-1}
    \label{fig5}
  \end{minipage}
  \hfill
  \begin{minipage}[b]{0.45\textwidth}
    \centering
    \includegraphics[width=\textwidth]{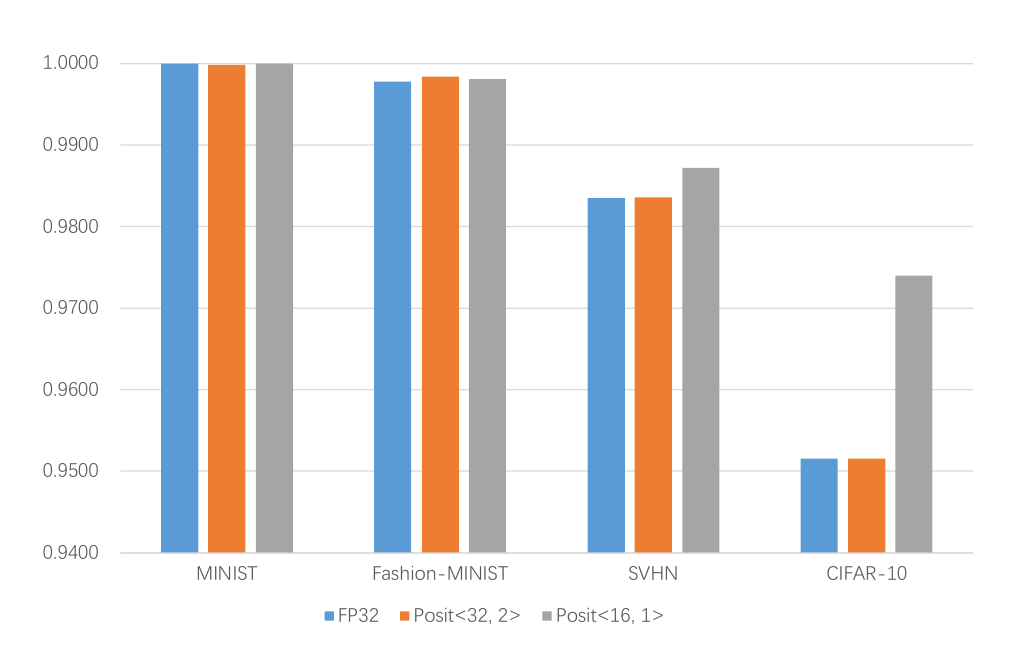}
    \caption{The accuracy of Posit-FP32 in TOP-5}
    \label{fig6}
  \end{minipage}
\end{figure}

  \begin{table}[h!]
    \centering
    \captionsetup{justification=centering}
    \caption{Hardware resources usage for different components}
    \begin{tabular}{|l|c|c|c|c|}
    \hline
     & LUTs & Muxes & DSPs & IOBs \\
    \hline
    PVU       & 46507 & 108  & 80  & 419 \\
    div       & 14830 & 68   & 80  & 0   \\
    mul       & 6479  & 8    & 0   & 0   \\
    dot-product & 6655 & 8    & 0   & 0   \\
    \hline
    \end{tabular}
    \label{table4}
  \end{table}

\section{Conclusion}
In this paper, we propose an open-source Posit Vector Arithmetic Unit (PVU) that can efficiently perform addition, subtraction, multiplication, division, and dot product operations in deep learning applications on low-power platforms. The PVU can be tightly integrated with RISC-V and, through inline assembly, map high-level languages to machine code. Additionally, a configurable generator for the PVU has been developed to support different Posit number systems in various low-power scenarios.

\section*{Acknowledgment}
This work is supported financially by Sichuan Natural Science Foundation for Distinguished Young Scholar (ID: 2023NSFSC1966),National Natural Science Foundation of China (ID: 61672438).


\begin{thebibliography}{24}

  \bibitem{1}
  Pullini A, Rossi D, Loi I, Tagliavini G, Benini L. Mr.Wolf: An Energy-Precision Scalable Parallel Ultra Low Power SoC for IoT Edge Processing[J]. IEEE Journal of Solid-State Circuits, 2019, 54(7): 1970-1981. DOI:10.1109/JSSC.2019.2912307.
  
  \bibitem{2}
  Wang S, Wang X, Xu Z, Chen B, Feng C, Wang Q, Ye T T. Optimizing CNN Computation Using RISC-V Custom Instruction Sets for Edge Platforms[J]. IEEE Transactions on Computers, 2024, 73(5): 1371-1382.
  
  \bibitem{3}
  C. Leong, “Softposit,” https://gitlab.com/cerlane/SoftPosit, 2018.
  
  \bibitem{4}
  Klöwer M, Düben PD, Palmer TN.SoftPosit.jl: A posit arithmetic emulator for Julia[EB/OL]. GitHub repository, 2020.
  
  \bibitem{5}
  S.D.Ciocirlan, D.Loghin, L.Ramapantulu, N.Tapus, and Y.M.Teo, “The Accuracy and Efficiency ofPosit Arithmetic,” 2021, arXiv:2109.08225.
  
  \bibitem{6}
  N.Shah, L.I.G.Olascoaga, S.Zhao, W.Meert, and M.Verhelst, “DPU:DAG processing unit for irregular graphs with precision-scalable posit
  arithmetic in 28 nm,” IEEE J. Solid State Circuits (JSSC), vol. 57, no. 8,
  pp. 2586-2596, 2022.
  
  \bibitem{7}
  N.Ho, D.T.Nguyen, H.D.Silva, J.L.Gustafson, W.Wong, and I.J.Chang, “Posit arithmetic for the training and deployment of generative
  adversarial networks,” in 2021 Design, Automation \& Test in Europe
  Conference \& Exhibition (DATE). IEEE, 2021, pp. 1350-1355.

  \bibitem{8}
  de Dinechin, F., Forget, L., Muller, J.-M., \& Uguen, Y. Posits: The Good, the Bad and the Ugly. Univ Lyon, INSA Lyon, Inria, CITI, Lyon, France.

  \bibitem{9}
  Carmichael, Z., Langroudi, H. F., Khazanov, C., Lillie, J., Gustafson, J. L., \& Kudithipudi, D. Deep Positron: A Deep Neural Network Using the Posit Number System. Neuromorphic AI Lab, Rochester Institute of Technology, NY, USA, National University of Singapore, Singapore.

  \bibitem{10}
  Cococcioni, M., Rossi, F., Ruffaldi, E., and Saponara, S., \textit{Fast Approximations of Activation Functions in Deep Neural Networks when using Posit Arithmetic}, Department of Information Engineering, Università di Pisa, Medical Microinstruments (MMI) S.p.A., 2020, Published: March 10, 2020.
  

  \bibitem{11}
  Murillo, R., Del Barrio, A. A., and Botella, G., \textit{Deep PeNSieve: A deep learning framework based on the posit number system}, \textit{Digital Signal Processing}, vol. 102, p. 102762, 2020, doi: 10.1016/j.dsp.2020.102762.
  
  \bibitem{12}
  Mallas, D., Murillo, R., Del Barrio, A. A., Botella, G., Piñuel, L., \& Prieto-Matias, M. PERCIVAL: Open-Source Posit RISC-V Core With Quire Capability[J]. IEEE Access, Vol. 10, No. 3, July-Sept. 2022.

  \bibitem{13}
  Mallasén, D., Del Barrio, A. A., \& Prieto-Matias, M. Big-PERCIVAL: Exploring the Native Use of 64-Bit Posit Arithmetic in Scientific Computing[EB/OL]. arXiv preprint, 2023, May.

  \bibitem{14}
  M. V. Arunkumar, S. G. Bhairathi, and H. G. Hayatnagarkar, “PERC: Posit Enhanced Rocket Chip,” in 4th Workshop on Computer Architecture Research with RISC-V (CARRV'20), 2020, p. 8.

  \bibitem{15}
  S. Tiwari, N. Gala, C. Rebeiro, and V. Kamakoti, “PERI: A Configurable Posit Enabled RISC-V Core,” ACM Transactions on Architecture and Code Optimization, vol. 18, no. 3, pp. 1–26, Jun. 2021.

  \bibitem{16}
  S. D. Ciocirlan, D. Loghin, L. Ramapantulu, N. Tapus, and Y. M. Teo, “The Accuracy and Efficiency of Posit Arithmetic,” 2021, arXiv:2109.08225.

  \bibitem{17}
  M. Cococcioni, F. Rossi, E. Ruffaldi, and S. Saponara, “A Lightweight Posit Processing Unit for RISC-V Processors in Deep Neural Network Applications,” IEEE Transactions on Emerging Topics in Computing, no. 01, pp. 1–1, Oct. 2021.

  \bibitem{18}
  Li, Q., Fang, C., and Wang, Z., \textit{PDPU: An Open-Source Posit Dot-Product Unit for Deep Learning Applications}, in \textit{2023 IEEE International Symposium on Circuits and Systems (ISCAS)}, 2023, IEEE.
  

  \bibitem{19}
  K. He, X. Zhang, S. Ren, and J. Sun, “Deep residual learning for image recognition,” in 2016 IEEE Conference on Computer Vision and Pattern Recognition (CVPR), 2016, pp. 770–778.

  \bibitem{20}
  Murillo, R., Del Barrio, A. A., and Botella, G., \textit{Deep PeNSieve: A Deep Learning Framework Based on the Posit Number System}, \textit{Digital Signal Processing}, vol. 102, p. 102762, 2020, doi: 10.1016/j.dsp.2020.102762.
  

  \bibitem{21}
  RISC-V Foundation, \textit{RISC-V Vector Extension Version 1.0}, 2021. Available at: \url{https://github.com/riscv/riscv-v-spec}, Accessed: 2025-02-26.
  

  \bibitem{22}
  C. L. and D. S. Patterson, \textit{Chisel: A Hardware Design Language for the Modern Era}, 2012. Available at: \url{https://www.chisel-lang.org/}.
  
  \bibitem{23}
  IEEE Standard 754-2008, \textit{IEEE Standard for Floating-Point Arithmetic}, IEEE, 2008, doi: 10.1109/IEEESTD.2008.4610935.
  
  \bibitem{24}
  J. L. Gustafson, \textit{Beating Floating Point at its Own Game: Posit Arithmetic}, \textit{IEEE Access}, vol. 7, pp. 11175-11183, 2019, doi: 10.1109/ACCESS.2019.2910693.


\end{thebibliography}
\end{document}